\documentclass[12pt]{iopart}

\newcommand{\text}[1]{\mathrm{#1}} 
\usepackage{cite} 

\expandafter\let\csname equation*\endcsname\relax
\expandafter\let\csname endequation*\endcsname\relax
\usepackage{amsmath}
\usepackage{iopams}
\usepackage{subcaption} 

\usepackage{amssymb} 
\usepackage{comment} 
 \usepackage{graphicx} 
 \usepackage{fix-cm} 

\usepackage{color}

\usepackage{soul}

\begin{document}
\newcommand\barparen[1]{\overset{(-)}{#1}}

\title[]{Searching for Long-Lived Particles in Free Neutron Experiments} 


\author{B. Meirose$^{1,2}$, R. Nieuwenhuis$^2$, R. Pasechnik$^2$, \\
H. Gisbert$^3$, L. Vale~Silva$^4$, \\
D. Milstead$^5$}
\address{$^1$ Institutionen f{\"o}r Fysik, Chalmers Tekniska H\"{o}gskola, Sweden}
\address{$^2$ Department of Physics, Lund University, Professorsgatan 1, 22363 Lund, Sweden}
\address{$^3$Escuela de Ciencias, Ingenier\'ia y Dise\~{n}o, Universidad Europea de Valencia, \\
Passeig de la Petxina 2, 46008 Valencia, Spain}
\address{$^4$ Departamento de Matem\'{a}ticas, F\'{i}sica y Ciencias Tecnol\'{o}gicas,\\
Universidad Cardenal Herrera-CEU, CEU Universities,\\
46115 Alfara del Patriarca, Val\`{e}ncia, Spain}
\address{$^5$ Department of Physics, Stockholm University, 106 91 Stockholm, Sweden}

\ead{bernhard.meirose@fysik.lu.se}
\vspace{10pt}
\begin{indented}
\item[] June 10, 2025
\end{indented}
\date{\today}

\begin{abstract}
We explore the decay of free neutrons into exotic long-lived particles, whose decays could be detected in the next-generation free neutron experiments. We show that such a possibility is viable as long as the exotic particle is highly mass-degenerate with the neutron, avoiding exclusion by large-volume detectors. We estimate the number of observable events and identify the most promising final states from both theoretical and experimental perspectives. Our analysis highlights the unique capability of the HIBEAM-NNBAR experiment at the European Spallation Source to probe this unexplored region of parameter space, opening a new avenue for exploring physics beyond the Standard Model. We estimate that several events per year could be observed in the NNBAR experiment.
\end{abstract}

\section{Introduction}
Neutrons are nearly ubiquitous constituents of observable matter and can be copiously produced in the laboratory. It is therefore possible to use ultra-rare processes involving neutrons as sensitive probes for feeble interactions that reveal physics beyond the Standard Model (SM).

The exploration of exotic rare neutron decays beyond the dominant $\beta$-decay mode has traditionally been conducted by large-volume experiments due to the enormous number of neutrons needed for high-precision measurements. However, it is often overlooked that large-mass detectors are blind to decays in certain kinematic regions.

Although free neutrons decay within minutes, bound neutrons are mostly stable because, in stable nuclei, the energy gained from the neutron decay is lower than the energy required to accommodate an additional proton in the nuclear core. This is a direct consequence of the near-degeneracy in mass between the proton and the neutron. If exotic particles that interact with and are close in mass to the neutron exist, they would have escaped detection by large-volume experiments. This argument was explored in Ref.~\cite{Fornal:2018eol} within a scenario involving a stable exotic fermion, which can be interpreted as a dark matter candidate. While decay to a stable particle may be the most natural hypothesis for an exotic neutron decay, the particle may, like the neutron itself, be quasi-stable, with a relatively long lifetime. This possibility requires investigation through searches using free neutrons.

In this paper, we propose fully exploring this gap about a particle as pivotal to our understanding of nature as the neutron, taking advantage of the intense flux of free neutrons that will be delivered by the European Spallation Source (ESS)~\cite{Garoby:2017vew}, currently under construction in Lund, Sweden. In particular, we discuss the potential of the future HIBEAM-NNBAR experiment~\cite{Addazi:2020nlz,Santoro:2023izd,Santoro:2024lvc} at the ESS to measure these exotic decays, as it is the only free-neutron experiment program with the required detector capabilities to accurately reconstruct the final states of interest.
Clearly, the ESS has a potential impact in physics that goes beyond the scope of the present paper, see e.g. Ref.~\cite{Abele:2022iml}.

The paper is divided as follows. \textcolor{black}{In Section~\ref{sec:hibeam_nnbar} we briefly present the HIBEAM-NNBAR experimental program.} In Section~\ref{const}, we discuss the existing experimental constraints on exotic neutron decays, both from free-neutron experiments and large-volume detectors. In Section~\ref{limits}, we discuss the limits on long-lived particles, while in Section~\ref{sec_min_width}, we examine the discovery potential of free-neutron experiments for exotic long-lived particles under different spin hypotheses. In Section~\ref{motivation} we discuss the motivation for searching for exotic neutron decays and briefly mention the theoretical scenarios within which these searches are embedded. In Section~\ref{theory}, we outline a general theoretical framework relevant to these searches, discussing final states of interest in Section~\ref{measure}, followed by a discussion of the experimental setup in Section~\ref{experiment}. We conclude in Section~\ref{conclusion}.

\section{Overview of the HIBEAM-NNBAR Program}
\label{sec:hibeam_nnbar}

The HIBEAM-NNBAR program~\cite{Addazi:2020nlz} at
ESS is a staged effort designed to investigate baryon number violation and other precision tests of fundamental symmetries using free neutrons. The program consists of two phases: the High-Intensity Baryon Extraction and Measurement (HIBEAM) as the first stage, followed by the second-stage NNBAR experiment. Both will operate at the ESS, leveraging its high-intensity pulsed neutron source.

HIBEAM~\cite{Santoro:2023izd} will be located at $L \approx 50$~m from the moderator and use a cold neutron flux of $\Phi_n \approx 10^{12} $ n/s. It will host a broad range of fundamental physics experiments, such as searches for neutron-antineutron oscillation ($n \rightarrow \bar{n}$), neutron-to-sterile neutron conversion ($n \rightarrow n'$), hadronic parity violation, axion-like particle searches by Ramsey interferometry, and a search for a non-zero electric charge of the neutron.

\begin{figure}[!htb]
	\centering
	\includegraphics*[width=0.63\textwidth]{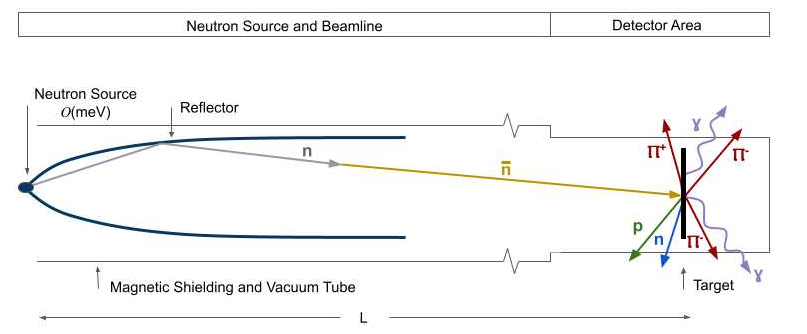}
	\captionsetup{width=0.93\linewidth}
	\caption{A schematic view of the NNBAR experiment. Taken from Ref.~\cite{Yiu:2022faw}.}
	\label{fig:nnbar_general}
\end{figure}

The NNBAR~\cite{Santoro:2024lvc} second stage will push the baseline to $L \approx 200$~m and benefit from much higher neutron fluxes up to \( \Phi_n \approx 10^{14} \) n/s, facilitated by the Large Beam Port (LBP), neutron mirror optics with supermirror coating $(m = 6)$, and a high-power 5~MW proton linac. Figure~\ref{fig:nnbar_general} 
shows a schematic view of the NNBAR experiment.

Together, the HIBEAM and NNBAR stages constitute a flexible and high-intensity platform for precision neutron experiments at the ESS. In the following sections, we primarily evaluate the potential sensitivity to rare neutron decay channels under the specific conditions of the NNBAR stage, while also discussing selected channels accessible at HIBEAM.

\section{Experimental constraints on neutron decays}
\label{const}

Experimental constraints on the branching ratios of decay modes of the neutron come from various experiments, with the most sensitive searches conducted by the Super-Kamiokande (SK) experiment~\cite{Abe:2013gga,Super-Kamiokande:2002weg,Takhistov:2016eqm}. For most channels, the experimental limits on the neutron lifetime exceed \(10^{34}\) years, which can, for instance, be easily translated into lower bounds on the energy scale suppressing higher-dimensional effective operators violating baryon and lepton numbers \cite{Weinberg:1979sa}. However, these limits only apply to bound neutrons, which are the only types available in large underground experiments. This distinction is crucial because any exotic decay of the neutron, where the total mass of the final-state particles is very close to the neutron mass, would be highly suppressed. Such decays are promptly forbidden for neutrons bound in nuclei due to kinematic constraints.

To our knowledge, this argument was first explored in Ref.~\cite{Fornal:2018eol} to investigate the possibility of invisible neutron decays as an explanation for the discrepancy between in-beam and in-bottle neutron lifetime measurements. New data from free-neutron experiments~\cite{Czarnecki:2018okw, Dubbers:2018kgh} have excluded dark neutron decays as an explanation for the neutron anomaly at \(4\,\sigma\) level. Furthermore, a recent study sets an upper bound on the dark neutron branching ratio (BR) of \(\mathcal{O}(10^{-5})\), based on the dark neutron decay model proposed in Ref.~\cite{Fornal:2018eol}, using a high-intensity \(^{6}\text{He}^+\) beam at GANIL~\cite{LeJoubioux:2023usk}. These experiments, however, have not excluded the possibility of exotic rare decays within the small mass gap region to which large underground experiments are blind. Free-neutron experiments do set limits on how large, relative to the known neutron branching ratio, exotic rare decays are allowed to be. However, even the most stringent constraints on exotic neutron decay branching ratios are in general no better than BR~\(\leq 0.1\%\)~\cite{Dubbers:2018kgh}, meaning that limits are set only at the per mille level.

The ultimate factors driving limits obtained from free-neutron experiments are the total \textcolor{black}{neutron flux} for which decays can be measured, integrated over the total observation time. The cold neutron intensity delivered by ESS to both HIBEAM and NNBAR will be higher than any existing neutron facility in the world~\cite{Santoro:2023izd,Santoro:2024lvc}.

Naturally, neutron flight time, as well as the duration of data-taking, also improves sensitivity to discovery. Here again, NNBAR has a significant edge over previous experiments. It will be the longest free cold neutron beam experiment with a flight length extending to $\sim$ 200~m~\cite{Santoro:2023izd}. \textcolor{black}{ Considering the neutron flux, velocity distribution, and total data-taking time, this corresponds to an integrated exposure of approximately $10^{18}$--$10^{19}$ neutron-lifetimes.} In comparison, the total distance from the cold neutron source to the detector in Ref.~\cite{Dubbers:2018kgh} was well below 100~m.

An important remark is that our assumption regarding the exclusion of an exotic signal by any free-neutron lifetime experiment is conservative. To the best of our knowledge, no neutron lifetime experiment has actively searched for the \textit{visible} final states of interest following the decay of an exotic particle produced in the neutron decay, nor have selection criteria been developed. Furthermore, the searches require full reconstruction of the neutron decay products with the aid of particle trackers and electromagnetic and hadronic calorimeters, which are not typically used in neutron lifetime experiments. However, the exclusion assumption is acceptable, assuming that all observed decays are accounted for by standard neutron beta decays within experimental uncertainties, which is the reason to assume the limit BR~\(\leq 0.1\%\)~\cite{Dubbers:2018kgh} is valid in our case.

Finally, note that in the discussion of the beam-bottle neutron lifetime anomaly~\cite{Paul:2009md}, the typical exotic branching ratio is ten times larger. For the small exotic branching ratios under discussion in this paper, we do not expect the exotic decay channel to affect substantially the dynamics of neutron stars; see Ref.~\cite{Fornal:2023wji} and references therein.

\section{Limits on long-lived particle production from neutron decays}
\label{limits}

To provide a concrete example,
consider first the process $n \rightarrow X + \gamma$, where $X$ is a fermionic particle with a mass close to that of the neutron (the final state photon energy \(0.782 \, \text{MeV} <E_\gamma\leq 1.665 \, \text{MeV} \), following Ref.~\cite{Fornal:2018eol}). To be consistent with nuclear stability constraints, particularly measurements from Super-Kamiokande, the particle $X$ must either be completely stable or long-lived. If the width of $X$ is sufficiently large (indicating a short lifetime), a significant rate of events would be expected in Super-Kamiokande due to off-shell decays. In this case, the small gap hypothesis would no longer hold, as the neutron would ultimately decay into visible states—a phenomenon not observed.

The key question is: what constitutes a `sufficiently large' width for $X$? Specifically, what is the maximum width of $X$ consistent with the absence of a statistically significant signal at Super-Kamiokande? 
The probability that a particle with invariant mass \(M_X\) is off-shell—assuming its on-shell mass is given by 
\(M_X^\text{on-shell} = M_n - m_\text{gap}\), where \(M_n = 939.565 \, \text{MeV}\) is the mass of the neutron and \textcolor{black}{we choose \(m_\text{gap} \leq 1 \, \text{MeV}\), although the need for a smaller range based on experimental reasons will be motivated below}—and \(M_X\) falls within the range 
\(0 \leq M_X \leq 937.9 \, \text{MeV}\), is given by the integral of the Breit-Wigner distribution. Here, the upper limit \(937.9 \, \text{MeV}\) is determined as \(M_n - m_{\rm upper}\), with $ m_{\rm upper} = 1.665 \, \text{MeV} $, \textcolor{black}{such that the kinematical configuration of the decay becomes accessible in bound neutron experiments when $M_X$ goes below this upper limit \cite{Fornal:2018eol}, which is only possible due to the finite, non-vanishing total width $\Gamma_X$ of $X$}. The expression for the probability is:

\begin{equation}
\label{BWigner}
P(X_\textup{off-shell}) = \int_{0}^{937.9 \, \text{MeV}} \frac{1}{\pi} \cdot \frac{\Gamma_X}{(M_X - M_X^\text{on-shell} \, \text{})^2 + \frac{\Gamma_X^2}{4}} \, dM_X \,.
\end{equation}
\textcolor{black}{In considering this upper limit of the integration while discussing SK bounds, we opt for being conservative. Indeed, the value of $m_{\rm upper}$ given above is needed in order to avoid the bound set when opening the transition ${}^{9}{\rm Be} \to {}^{8}{\rm Be}$ (whose mass difference is $937.9$~MeV), while in addressing SK bounds discussed below it is sufficient to avoid the transition ${}^{16}{\rm O} \to {}^{15}{\rm O}$ at large rates (of mass difference $923.9$~MeV).
Hereafter, we refer to the $X$'s on-shell mass simply as $M_X$.}

The rate of events, under the assumption of nuclear suppression, is calculated as:
\begin{equation}
    \Gamma = \Gamma_n \cdot \text{$\text{BR}$} \cdot P(X_\textup{off-shell}),
\end{equation}
where $\text{BR} = 10^{-3}$, is the branching ratio of the neutron to the exotic decay $X$ and $\Gamma_n = 1/878$ Hz is the neutron width~\cite{ParticleDataGroup:2018ovx}. Data collected by the Super-Kamiokande water Cherenkov experiment between 1996 and 2018~\cite{Super-Kamiokande:2024qbv} corresponds to a combined exposure of 370 kiloton$\cdot$yrs. This equates to approximately $10^{35}$ neutrons over one year of data taking. The total number of expected observed decays over a year is thus:
\begin{equation}
    \label{eq:N_decays}
    N_{\text{decays}} = N_{\text{SK}} \cdot \Gamma \cdot T_{\text{year}} \cdot \epsilon,
\end{equation}
where $N_{\text{SK}} = 10^{35}$ is the number of neutrons, $T_\text{year}$ $\sim$ $3.2 \times 10^{7}$~s and $\epsilon$ is Super-Kamiokande's efficiency to detect such a decay .

The  $X$ decays for which the Super-Kamiokande experiment would have been least efficient is when $X$ is a fermionic particle. In this case, there are no two-body decays of the $X$ particle to charged lepton pairs, which are efficiently reconstructed by the detector. Instead, all two-body decays involve either a hadron, a neutrino, or both (see later discussion in Section~\ref{measure}). Indeed, the latest reported efficiency for reconstructing neutron decay channels with a lepton (electron or muon) plus a charged pion was approximately 12\%~\cite{Super-Kamiokande:2017gev}.

Moreover, the average efficiency for reconstructing channels with a $\rho$ meson and a charged lepton was around 1\%, primarily due to the difficulty in detecting charged pions in a large water Cherenkov detector. Therefore, to account for various factors, we consider three benchmark scenarios, with efficiencies $\epsilon = 5\%, 10\%, \text{ and } 20\%$, and background expectations of 1, 0.6, and 0.3 events, respectively. The three benchmark scenarios are labeled, ``loose", ``medium" and ``tight", respectively. \textcolor{black}{These benchmark scenarios encompass efficiencies ranging from the previously discussed neutron decay channel with the highest reported efficiency—lepton plus a charged pion (12\%)~\cite{Super-Kamiokande:2017gev}—which we conservatively take as 20\%, motivated by an earlier result for this channel that reported an efficiency of 19\%~\cite{Super-Kamiokande:2012ngt}. The corresponding background expectation was reported as $0.41 \pm 0.13$ events~\cite{Super-Kamiokande:2017gev}, justifying our conservative assumption of 0.3 background events for the ``tight'' scenario. At the other end of the spectrum, we consider channels involving a $\rho$ meson and a charged lepton, which have the lowest reported efficiency (around 1\%) and an associated background expectation of $0.96 \pm 0.28$ events~\cite{Super-Kamiokande:2017gev}. For our benchmark estimate, we conservatively assume a 5\% efficiency and round the background to 1 event. This conservative treatment is adopted because these channels serve as proxies for the experimental efficiencies, had the corresponding decay modes occurred at appreciable rates in Super-Kamiokande. For the medium scenario, we assume that some of the new channels may exhibit efficiencies and background expectations intermediate between the benchmark cases, due to differences in kinematics and detector response.}

If $X$ is a boson, however, there are clean decays to lepton pairs, and the expected efficiency would be much higher, potentially as high as 50\% as inferred from Super-Kamiokande's dinucleon searches~\cite{Super-Kamiokande:2018apg}, and virtually background-free. In principle, the suppression factor $P(X_{\text{off-shell}})$ could be made arbitrarily strong (i.e., a very small number) to compensate for any high efficiency, but this would also make it unreachable for reconstruction in future experiments with free neutrons. While this might not be the most common scenario, one could still hypothesize that if $X$ is a boson, it could be completely leptophobic. While requiring leptonic decays to be effectively forbidden is a less general model, it is not ruled out. Indeed, nothing prevents $X$ from being a particle similar to an exotic long-lived $\rho^0$ meson, with decay modes restricted to a pair of charged pions. Under this assumption, the benchmark scenarios for a fermionic $X$ apply equally.

To assess the statistical significance of Super-Kamiokande detecting a signal above the background, a threshold for a \(3\,\sigma\) deviation was computed using the Poisson cumulative distribution function (CDF). The threshold \( N_{3\,\sigma} \) is defined as the smallest integer satisfying \( P(N \leq N_{3\,\sigma} \mid B_{\text{obs}}) \geq 0.99865 \), where \( P(N \leq x \mid B_{\text{obs}}) \) is the CDF of a Poisson-distributed variable with mean \( B_{\text{obs}} \), which represents the expected number of background events, evaluated at a threshold 
$x$, which corresponds to a given statistical confidence level. The value \( 0.99865 \) corresponds to the probability of a standard normal variable exceeding \( 3\,\sigma \), given by \( P(Z \leq 3) = \frac{1}{\sqrt{2\pi}} \int_{-\infty}^{3} e^{-z^2/2} dz = 0.99865 \). The threshold \( N_{3\,\sigma} \) is determined numerically from this condition.

To estimate the probability of Super-Kamiokande having observed a statistically significant excess, a statistical simulation with $10^5$ independent trials was conducted. In each trial, the number of observed events was drawn from a Poisson distribution with mean \( N_{\text{obs}} + B_{\text{obs}}\). The number of trials in which the observed number of events exceeded the \(3\,\sigma\) threshold was recorded as \(\text{Count}_{3\,\sigma}\). The corresponding probability was then computed as \(P_{3\,\sigma} = \text{Count}_{3\,\sigma} / N_{\text{trials}}\). The requirement for Super-Kamiokande to have missed the signal is that the probability of observing a \(3\,\sigma\) excess was below 0.1, i.e., \(P_{3\,\sigma} < 0.1\). In other words, in 90\% of the pseudo-experiments, the reach of Super-Kamiokande would have been below the level of evidence, taking into account statistical fluctuations modeled by a Poisson distribution.

To meet these criteria for the three aforementioned benchmark scenarios we obtain:
\begin{align}
P(X_\textup{off-shell})_{l} &= 8 \times 10^{-36}~, \\ 
P(X_\textup{off-shell})_{m} &= 3 \times 10^{-36}~, \\ 
P(X_\textup{off-shell})_{t} &= 1 \times 10^{-36}~, 
\end{align}
where the indices represent the loose, medium and tight benchmark scenarios. Substituting these thresholds into Eq.~\eqref{BWigner}, \textcolor{black}{for \(m_\text{gap} \lesssim 20 \, \text{keV}\)
yields:}
\begin{align}
\Gamma_{X_l} &\lesssim 4.2 \times 10^{-35} \, \text{MeV}; \quad\tau_{X_l} \approx 5.0 \times 10^{5} \, \text{yrs}~, \\ 
\Gamma_{X_m} &\lesssim 1.7 \times 10^{-35} \, \text{MeV}; \quad\tau_{X_m} \approx 1.2 \times 10^{6} \, \text{yrs}~, \\ 
\Gamma_{X_t} &\lesssim 5.9 \times 10^{-36} \, \text{MeV};\quad \tau_{X_t} \approx 3.6 \times 10^{6} \, \text{yrs}~. 
\end{align}
These values are nearly thirty orders of magnitude shorter than typical lifetimes set by Super-Kamiokande ($\sim$ $10^{34}$ years) for neutrons decaying into exotic decays. 
\textcolor{black}{
Given the relatively long lifetime assumed for the \( X \) particle—of order \(10^5\) to \(10^6\) years—collider experiments and those designed to search for long-lived particles are not ideal for probing its production and decay. Displaced vertex searches at the LHC (ATLAS~\cite{ATLAS:2018tup}, CMS~\cite{CMS:2019qjk}) are sensitive to long-lived particle lifetimes ranging from picoseconds to tens of nanoseconds. Future detectors like MATHUSLA~\cite{MATHUSLA:2019qpy} extend this sensitivity to lifetimes of approximately 0.1 microseconds to milliseconds, while SHiP~\cite{SHiP:2021nfo} covers a similar range, from nanoseconds to microseconds. The much longer lifetime of the \( X \) particle places it well outside the sensitivity range of these collider-based searches. Moreover, due to the extremely small couplings involved, the production cross sections at colliders are negligible, rendering such long-lived particles effectively invisible—even in missing energy channels.
}

\textcolor{black}{
On the other hand, $X$'s lifetime is still many orders of magnitude smaller than the age of the universe, which excludes it as a viable dark matter candidate. Any primordial population would have long decayed, and even in the case of residual or rare astrophysical production, the couplings required for exotic neutron decay (\(g \sim 10^{-12}\) for $ 0.5 \lesssim M_n - M_X \lesssim 20 $~keV) imply very small elastic scattering cross sections with nuclei.
Thus, the low-background, high-intensity nature of cold neutron experiments offers a unique window into ultra-weakly coupled sectors and baryon number-violating processes that are inaccessible to other experimental programs.}

\textcolor{black}{
Finally, let us mention that proton decay via virtual neutrons $p \to n^\ast + e^+ + \nu_e$ is suppressed: given that $M_p-m_e = 937.8$~MeV and $M_X = M_n - m_{\rm gap} = 938.6$~MeV for $m_{\rm gap} = 1$~MeV, we meet again tiny probabilities $P(X_\textup{off-shell})$ of the $X$ particle decaying off-shell, needed in order to avoid the strong bounds from SK on neutron partial lifetimes; furthermore, a five-body phase-space suppression factor (or, alternatively, a loop suppression factor in some cases) applies.
}

\section{Discovery potential in free neutron experiments}
\label{sec_min_width}
The immediate question is, what is the minimum width (maximum lifetime) free neutron experiments can actually measure. Assume free neutrons fly a distance $L$ before they reach a detector, in which they have a distance $d$ (detector tube) to decay. The flux of neutrons decaying into $X$ receives contributions both from neutrons decaying into $X$ before entering the tube which surrounds the detector, as well as contributions from $X$ particles formed inside the tube. However, due to the much longer flight path, the flux of $X$ particles formed from neutron decays before reaching the tube is the dominant contribution and is given by:
\begin{equation}
\Phi_X = \Phi_n \cdot \left( 1 - e^{-t / \tau_{nX}} \right),
\end{equation}
where \( \tau_{nX}= 8.78 \times 10^{5} \, \mathrm{s} \) is the effective mean lifetime of the neutron due to its decay into $X$, and \(  t_{\text{}} = \frac{L_{\text{}}}{v}  \), where $L$ is the distance traveled before reaching the tube and $v$ is the velocity of the neutron. For NNBAR, $L \approx 200$~m \cite{Santoro:2023jly, Santoro:2024lvc}, and we estimate \( \Phi_n = 2 \times 10^{14} \)~n/s. The HighNESS moderator intensity~\cite{Santoro:2023jly} was used as input to estimate the brilliance transfer for each data point, which was subsequently summed to obtain the total brilliance. The calculations account for a neutron mirror optic (NMO) with a supermirror coating of $m = 6$ and include an overall loss factor of 0.4, corresponding to two reflections. This high flux is achievable due to the LBP,
which is three times the size of standard ESS beamports and was calculated for a 5~MW proton linac power.
For neutron velocities, we use a Maxwell-Boltzmann distribution for a liquid deuterium moderator at 20 K, which is representative of the conditions in the NNBAR experiment. The velocity considered ranges from 200 to 1000 m/s, capturing the typical cold neutron spectrum for this setup. Substituting these values, we arrive at the flux \( \Phi_X = 1.1 \times 10^{8} \) X/s. 

The number of $X$ particles decaying inside the tube is then given by:

\begin{equation}
N_{\text{total}} = \Phi_X \cdot ( 1 - e^{-t_\text{tube} / \tau_{X}}) \cdot T
\end{equation}
where \( T = 2~\mathrm{yrs} \), for 3 years of data assuming data is collected for two-thirds of a calendar year, $\tau_{X}$
is $X$'s lifetime for the three considered benchmark scenarios, and \( t_{\text{tube}} = \frac{d_{\text{tube}}}{v_{X}}  \), where $d_{tube}$ is the distance traveled by $X$ inside the detector tube, and $v_{X}$ is the velocity of the $X$ particle. We take $d_{\text{tube}} = 6$~m~\cite{Yiu:2022faw}, and $v_{X}$ is determined purely from the kinematics of the $ n \rightarrow X + \gamma$ decay, being highly dependent on \(m_\text{gap}\) as defined in Section~\ref{limits}. Since the energy released on the decay of the neutron determines the velocity of $X$, it strongly constrains the available time the particle flies inside the detector tube. We consider \(m_\text{gap} = 1 \, \text{keV} \) as the best theoretically motivated scenario (see Section~\ref{motivationGeV}) and with the best experimental reach, given the velocity dependence. Taking these values into account, the number of expected $X$ decays inside the detector during three years of data-taking for the NNBAR experiment is $N_{\text{total}} \approx 1$, $3$ and $7$ events, for the tight, medium and loose benchmark scenarios, respectively.
For a background-free selection and 50\% detector efficiency, both the medium and loose scenarios seem feasible, although dedicated simulations would be needed to confirm this. 

\subsection{HIBEAM}
Given that the prospects for detecting exotic neutron decays in the NNBAR experiment are already highly constrained, it may seem unlikely that the HIBEAM detector—operating with approximately two orders of magnitude lower neutron flux and a baseline of \(L \approx 50\) m—could identify any exotic neutron decay channels. However, an opportunity arises when considering final states that large water Cherenkov detectors struggle to distinguish from background events, where the effective detection efficiency \(\epsilon\) in Eq.~\eqref{eq:N_decays} is expected to be very low.  

Two particularly promising final states are \(X \to \nu + \gamma\) and \(X \to \pi^0 + \gamma\). Super-Kamiokande faces significant challenges in reconstructing photon-rich final states due to its limited capability for accurate vertex reconstruction and poor efficiency in detecting single photons without associated charged particles. The \(X \to \nu + \gamma\) channel is particularly problematic: since the neutrino is invisible and single-photon detection is highly inefficient, such events are virtually impossible for Super-Kamiokande to reconstruct. Additionally, substantial cosmic-ray and atmospheric neutrino backgrounds make event isolation even more difficult. In contrast, the HIBEAM detector offers superior event vertex reconstruction and momentum imbalance detection, making it particularly well-suited to identifying such signatures.

\subsection{Future measurements}
The reach of the NNBAR experiment concerning the mass gap between the neutron and its decay product, $X$, is limited to only a few keV (1–5 keV) due to the velocity of $X$. Beyond that, the expected number of decays inside the detector is too low, considering the constraints from Super-Kamiokande discussed in Section~\ref{limits}. It is tempting to imagine that the full $\sim 1.7$ MeV gap could be explored with a higher flux, but this is only true to a limited extent. Even with a tenfold increase in flux at NNBAR, only mass gaps up to $\sim 20$ keV could be probed. Beyond that, further flux increases would not yield more events, as no decays would have enough time to occur within the detector’s $6$ m size to be measured. Instead, exploring larger mass gaps would require a detector with a much longer flight path, which does not seem feasible in the foreseeable future. In this regard, the NNBAR experiment is already near the frontier of mensurability.

\section{Motivation for exotic nucleon decays}

\label{motivation}
Baryon and lepton number conservation are accidental symmetries of the SM, which are typically broken in its extensions. Actually, even within the SM itself, both lepton and baryon number are already violated through instanton effects~\cite{tHooft:1976rip}, although at highly suppressed and therefore non-observable rates at low energies. Baryon number violation is a key ingredient in explaining dynamically the baryon-antibaryon asymmetry of the Universe~\cite{Sakharov:1967dj}, a process termed baryogenesis, while in leptogenesis scenarios~\cite{Fukugita:1986hr, Davidson:2008bu} both lepton and baryon number violation are required, but the latter indirectly from the former through sphaleron processes. Exotic nucleon decays typically violate both lepton as well as baryon numbers, being a crucial prediction of GUTs through proton decay, although its importance extends far beyond them. Many theoretical extensions of the SM predict nucleon decays, including Supersymmetry~\cite{Nath:2006ut}, gauged \textit{B-L}~\cite{FileviezPerez:2010gw} as well as string-inspired intersecting D-brane models~\cite{Klebanov:2003my}.
Baryon number violation in exotic neutron decays, without proton decay, could have important consequences for baryogenesis, even though the exact influence on the baryon asymmetry of the Universe depends upon the exact underlying ultra-violet mechanism. For example, whether the process is thermally enhanced or thermally suppressed.
The question of a blind kinematic spot on exotic neutron decay searches, however small, is therefore an extremely important one. Even disregarding this important connection, from a purely experimental perspective, an open gap in such an important measurement is one that begs to be closed. Taking into account that the requirements for such decays to still be allowed are rather loose—the mass proximity of an exotic ``daughter" with the parent neutron and the requirement for it to be either stable or long-lived—a dedicated experimental effort to explore final-states which could have evaded detection needs to happen.

Moreover, the presence of nearly mass-degenerate exotic states is not unprecedented in physics. Similar patterns appear in neutral meson mixing,

neutrino oscillations, and the proton-neutron mass difference itself, which arises due to subtle effects such as electromagnetic corrections and quark mass differences. If the neutron has an exotic decay channel into a long-lived state, the required small mass gap could be the result of an underlying hidden-sector symmetry, analogous to the near-degeneracy observed in other systems.

\textcolor{black}{The requirement that the mass of the exotic particle $X$ be nearly degenerate with the neutron is necessary for the decay to remain viable under current experimental constraints. However, the most theoretically motivated scenario may in fact correspond to the limiting case $M_X = M_n$, featuring exact degeneracy, for instance if an additional symmetry or stabilization mechanism is present. In other words, while the experimental limits motivate fine-tuning in the theory, we argue that the resulting parameter choice may also naturally arise from well-motivated theoretical considerations.}

\subsection{Motivations for GeV particles at various spin hypotheses}
\label{motivationGeV}

Although we do not explore the phenomenological relation of the $X$ particle with dark matter, this is a viable possibility in principle.
A compelling theoretical motivation for searching for exotic \(\sim\) 1 GeV particles in neutron decays is that they naturally fit within dark portal scenarios. These frameworks propose the existence of a dark or hidden sector that interacts weakly or indirectly with the SM, often as a natural extension of the particle dark matter hypothesis. While dark matter itself might be a weakly interacting massive particle (WIMP) near the weak scale, the mass range of other particles within the dark sector remains largely unconstrained.

Theories suggesting hidden sector particles at MeV to GeV scales often predict highly suppressed couplings to the SM. This suppression allows mediators to acquire long lifetimes, enabling them to evade detection in high-energy collider experiments. Such particles offer distinctive experimental opportunities, particularly in searches for displaced vertices at facilities like the LHC~\cite{ATLAS:2024ocv} or in rare decay processes investigated by experiments like HIBEAM-NNBAR.

From a phenomenological perspective, long-lived exotic particles with masses around 1 GeV
emerge naturally in various theoretical models.

Spin-1 mediators, such as \(Z'\) bosons or dark photons, are particularly well-suited to dark portal models. These particles often arise in association with hidden \(U(1)\) gauge symmetries that kinetically mix with the SM photon~\cite{Holdom:1985ag, Arkani-Hamed:2008hhe}. Their long lifetimes result from suppressed decay widths, which can stem from small kinetic mixing parameters, large mediator masses, or restricted phase space for decays. Extensive research has explored vector mediators in the sub-GeV to few-GeV range, motivated by their potential to explain astrophysical phenomena like self-interacting dark matter and the structure of dark matter halos~\cite{Tulin:2013teo}. Similarly, secluded dark matter models propose hidden mediators with masses around 1 GeV, connecting the dark matter sector with the visible SM sector~\cite{Pospelov:2007mp}.

Besides the spin-1 scenario, other hypotheses are based on spin-0 or spin-\(\frac{1}{2}\) particles. For spin-0 particles, pseudo-Nambu-Goldstone bosons may appear due to the spontaneous breaking of approximate global symmetries, with their decays suppressed by the scale of the symmetry breaking. Axion-like particles (ALPs) are particularly interesting in this context. ALPs can arise in various extensions of the SM, often linked to solutions of the strong CP problem or other high-energy symmetries. Their coupling to SM fields is typically suppressed by a high-energy scale, resulting in long lifetimes and thus suppressed decay widths. Another possibility is scalar singlets in Higgs portal models, where either the interactions are suppressed by higher-dimensional operators or by forbidden decay channels.

Spin-\(\frac{1}{2}\) particles are also promising candidates. For instance, quasi-stable neutralinos in supersymmetric models can decay through weak-scale R-parity-violating processes or via suppressed mixing angles, naturally achieving long lifetimes in many scenarios. Another example is sterile neutrinos, which arise in seesaw mechanisms to explain known neutrino masses. These particles may couple very weakly to SM neutrinos, as small coupling constants lead to small decay widths and thus long lifetimes.

\section{Theoretical framework}
\label{theory}
The interaction between a neutron and a new exotic particle can be described by different Lorentz-invariant Lagrangians, depending on the nature of the new particle. We use effective Lagrangians to describe each interaction since at the kinetic energies of free neutron experiments like HIBEAM-NNBAR (meV), it is appropriate to treat the neutron as a fundamental particle. Furthermore, the main goal of the present work is to discuss the possibility of these decays being allowed from a broad perspective without assuming any particular underlying microscopic model at the quark level.

For scalar
and vector hypotheses, the coupling of a nearly mass-degenerate particle with a neutron involving a left-handed neutrino ($\nu_L$) allows for a two-body decay, while for a fermionic particle, coupling with a photon would allow for the maximum degree of degeneracy.
In the former case, since SM neutrinos are left-handed, parity and C-parity are maximally violated.

For the sake of definiteness, we define the following interactions mediating exotic neutron decay:

\begin{equation}
    \mathcal{L}_{\text{spin-0}} = g \, \bar{n} \,\nu_L \,X + \text{h.c.}~,
    \label{eq:scalar}
\end{equation}
\vspace{-9mm}

\begin{equation}
    \mathcal{L}_{\text{spin-1}} = g \, \bar{n}\, \gamma^\mu\, \nu_L\, X_\mu + \text{h.c.}~,
    \label{eq:vector}
\end{equation}
\vspace{-9mm}

\begin{equation}
    \mathcal{L}_{\text{spin-1/2}} = \frac{g'}{\Lambda} \, \bar{n}\, \sigma^{\mu\nu}\, (F + i \,G\, \gamma_5)\, X\, F_{\mu\nu} + \text{h.c.}~.
    \label{eq:fermion}
\end{equation}
\vspace{+1mm} 

In the first two cases (equations~\eqref{eq:scalar} and \eqref{eq:vector}) a single dimensionless coupling constant \( g \) is necessary, while a minimal coupling to the electromagnetic field through the covariant derivative
is not possible
since both fermions are electrically neutral.
Here, \( F_{\mu\nu} \) denotes the electromagnetic field strength tensor, and \( \sigma^{\mu\nu} = \frac{i}{2} [\gamma^\mu, \gamma^\nu] \) is the antisymmetric combination of gamma matrices, while \( \Lambda \) represents the energy scale at which new heavy degrees of freedom start manifesting.
Thus, for the fermionic case, it is assumed neutrons interact through a magnetic $F$ (or electric $G$) dipole moment, which couples to the magnetic (or electric, respectively) field part of the electromagnetic field strength tensor (obviously, $g'$ could be absorbed into the definitions of the coupling constants $F$ and $G$; in the following, the latter are assumed to be of $\mathcal{O} (1)$).

These equations describe the neutron's decay into the exotic state $X$, while the decays of $X$ into two fermions (\(f\)) are governed by standard
Lagrangians:
\begin{equation}
    \mathcal{L}_{X_\text{spin-0}} = g_X\, X \,\bar{f}\, (y + z\, \gamma_5 )\, f + \text{h.c.}~, \quad 
    \mathcal{L}_{X_\text{spin-1}} = g_X\, X_\mu \,\bar{f} \,\gamma^\mu\, (v + a \,\gamma_5 ) \,f + \text{h.c.}~.
\end{equation}
These terms follow from the spin hypothesis of \(X\) (as before, $g_X$ could be absorbed into the definitions of $y$ and $z$, or $v$ and $a$). Similar considerations apply to the decay of \(X\) into bosonic final states (e.g., mesons, or photons), which can be treated analogously
by requiring Lorentz invariance, and considering the relevant degrees of freedom at the low energy scale.
{\color{black} It is implicitly assumed that $X$ carries no electric charge, nor QCD color. On the other hand, a weak-isospin charge of $X$ is disfavored, e.g., to avoid Drell-Yann constraints on the production of the charged particles that would be found in the same multiplet as $X$, see e.g. Ref.~\cite{Cirelli:2005uq}.}

\textcolor{black}{We also note that, without committing to a particular ultraviolet model of the interaction $n X$ one might ask whether it could have any observable effects on other meson or baryon decays. We emphasize that this is not necessarily the case. The decay may be accompanied by the isospin-breaking first-generation quark combination (\( udd \)), with a suppressed coupling strength to other flavor combinations. In such a case, a decay into heavier baryons would be extremely small. The existence of the coupling $n X$ then, does not necessarily indicate visible signals in other hadronic systems.}

The coupling constant \(g_X\) of $X$ to its decay products is assumed to be universal across all decay channels of \(X\), consistent with certain dark-sector models.

This framework is consistent with the dark portal motivation discussed in Section~\ref{motivationGeV}. Exotic particles with suppressed couplings and extended lifetimes naturally align with dark portal scenarios, and the universality of \(g_X\) captures the phenomenological essence of many dark-sector models.

\begin{figure}[ht]
    \centering
    \includegraphics[width=0.8\textwidth,trim={0cm 0.15cm 0cm 0cm},clip]{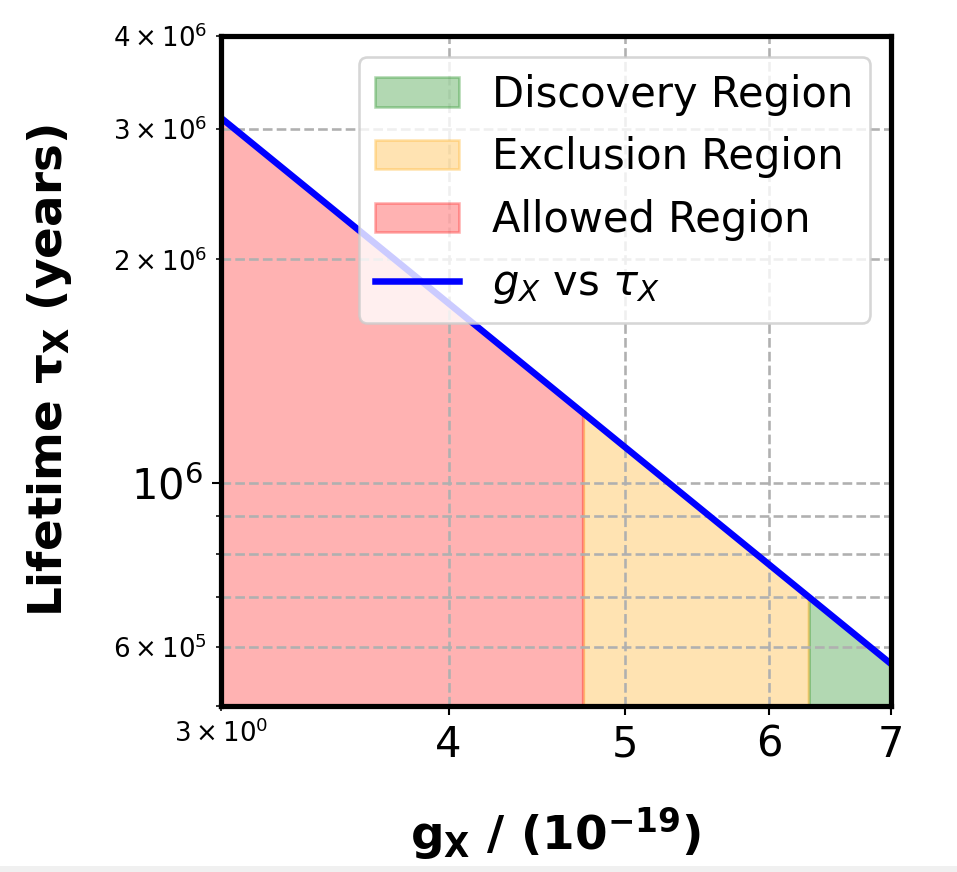}
    \caption{The decay lifetime $\tau_X$ as a function of the coupling strength $g_X$. The green, orange, and red shaded regions represent the discovery (5 or more events), exclusion (at least 3 events), and allowed (fewer than 3 events) regions, respectively.}
    \label{ggXtau}
\end{figure}

Up to order-one factors dependent on the specific model, the decay width of the exotic particle is given by 
$\Gamma_X \sim \frac{M_X}{4\pi} g_X^2$, while the decay width of the neutron to the exotic state is given by 
$\Gamma_{n} \sim \frac{M_n - M_X}{4 \pi \cdot \text{BR}} g^2$, where \( M_n \) is the mass of the neutron, \( M_X \) is the mass of the exotic particle, and \( \text{BR} \) is the branching ratio of the exotic neutron decay. The maximum and minimum values of $\Gamma_X$ are taken from Section~\ref{limits} and are therefore within the allowed lifetimes, taking into account limits imposed by Super-Kamiokande. Solving for the couplings yields the following ranges: 
$g_X = 2.8 \times 10^{-19}$ to $7.5 \times 10^{-19}$, and the exotic neutron decay coupling 
$g = 3 \times 10^{-12}$ for $ M_n - M_X = 1 $~keV.

Figure~\ref{ggXtau} shows the relationship between the coupling strength $g_X$ and the lifetime $\tau_X$. The horizontal axis depicts the coupling strength $g_X$ scaled by $10^{-19}$, while the vertical axis shows the $X$ particle lifetime in years. {\color{black}If further decay channels of the $X$ particle are experimentally accessible with the same coupling constant $g_X$, then the black curve in Figure~\ref{ggXtau} translates downwards.} The shaded regions show the discovery region in green (more than 5 events), the orange region represents the exclusion zone (at least 3 events), and the red region denotes a region with less than 3 events expected.

\section{Measurable final states}
\label{measure}
Taking into account the small mass gap argument, the leading measurable final states are: $ n \rightarrow X + \nu$, for a vector or scalar $X$ particle, and $ n \rightarrow X + \gamma$ for a fermion. 
Regarding the spin hypothesis, a fermionic $X$ should easily be distinguishable from a vector/scalar $X$ for certain channels, since its spin can be inferred from the final-state products. Experimentally, it could be challenging to distinguish between a scalar and a vector particle, since, unlike general-purpose experiments that rely on statistics to detect a new resonance, the event rates could be too low. Indeed, as is the case of standard neutron-to-antineutron searches, even a single event---assuming a zero-background experiment---could be enough to constitute a discovery. Thus, accumulating a sufficient number of events to distinguish spin hypotheses via angular distributions would not be possible in such scenarios. 

Of course, here one assumes that nothing is in principle known about the C-parity of the new particle. Assuming C-parity in the interactions leading to its decay, if $X$ has \textit{positive} C-parity, then a decay into two photons is possible regardless of whether $X$ is a scalar or a vector. However, if $X$ is a \textit{vector} particle with \textit{negative} C-parity, then such a decay would be forbidden by C-parity conservation (as in the case of known vector mesons like $J/\psi$, according to Furry’s theorem). Thus, while a two-photon decay could be a strong indication that $X$ is a scalar, a vector interpretation remains viable if $X$ has positive C-parity.

\subsection{Two-body decay channels of the $X$ particle}
The following two-body decay channels are possible. Channels containing only neutrinos are excluded since they would leave no measurable signal. 

\begin{itemize}
    \item \textbf{Channels exclusive for fermionic X:}
    \begin{itemize}
        \item \( X \rightarrow P^\pm + e^\mp \), where $ P^\pm = \pi^\pm, \, K^\pm, \, \rho^\pm $
        \item \( X \rightarrow P^0 + \nu \), where $ P^0 = \pi^0, \, \overset{\textbf{\fontsize{2pt}{2pt}\selectfont(---)}}{K} {}^0, \, \eta, \, \omega, \, \gamma $
    \end{itemize}
\end{itemize}

We tacitly assume $X$ is not a Majorana particle, in order to avoid $ |\Delta L| = 2 $ processes.

\begin{itemize}
    \item \textbf{Channels exclusive for scalar/vector X:}
    \begin{itemize}
        \item \( X \rightarrow \pi^{+,\,0} + \pi^{-,\,0} \) 
        \item \( X \rightarrow \rho^{\pm,\,0} + \pi^{\mp,\,0} \) 
        \item \( X \rightarrow K^{\pm,\,0} + \pi^{\mp,\,0} \) 
        \item \( X \rightarrow \gamma + \gamma \) 
    \end{itemize}
\end{itemize}

It is worth noticing that while channels like \( X \rightarrow \pi^+ + \pi^- \) require, theoretically, a bosonic parent particle, experimentally, it might be difficult to disentangle these from fermionic three-body decays like \( X \rightarrow \pi^+ + \pi^- + \nu\) since any energy reconstruction inefficiency could be misidentified as energy not measured due to a low-energy neutrino, depending also on tracking resolution (two-body decays are perfectly back-to-back), especially taking into account that no high-statistics distributions are likely to be possible.

\section{Experimental setup for exotic decays}
\label{experiment}

\begin{figure}[h]
  \centering
  \begin{subfigure}[b]{0.40\textwidth}
    \centering
    \includegraphics[width=\textwidth]{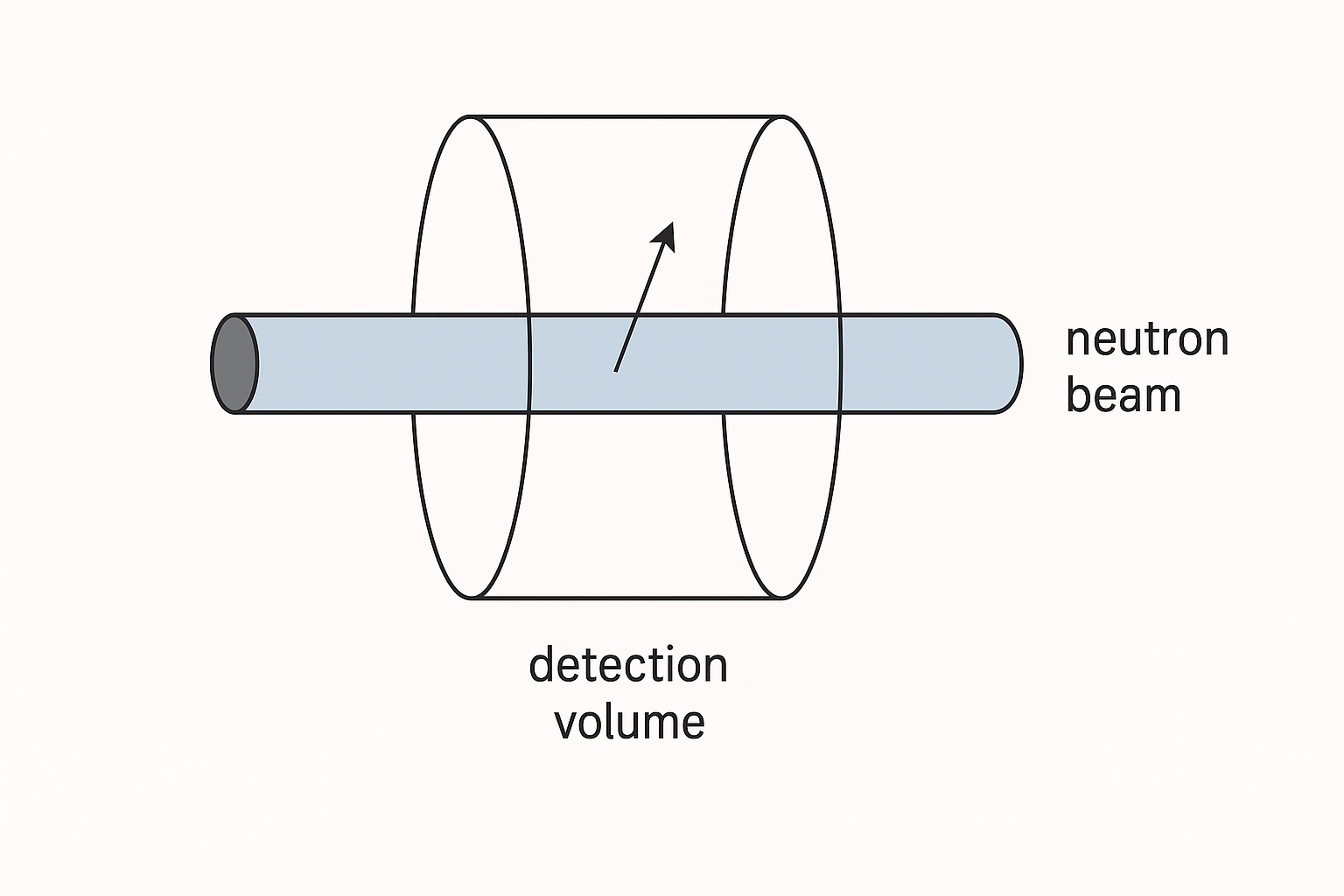}
    \caption{Neutron decaying inside the beam pipe surrounded by a detector.}
    \label{fig:rare_neutrons}
  \end{subfigure}
  \hfill
  \begin{subfigure}[b]{0.55\textwidth}
    \centering
    \includegraphics[width=\textwidth]{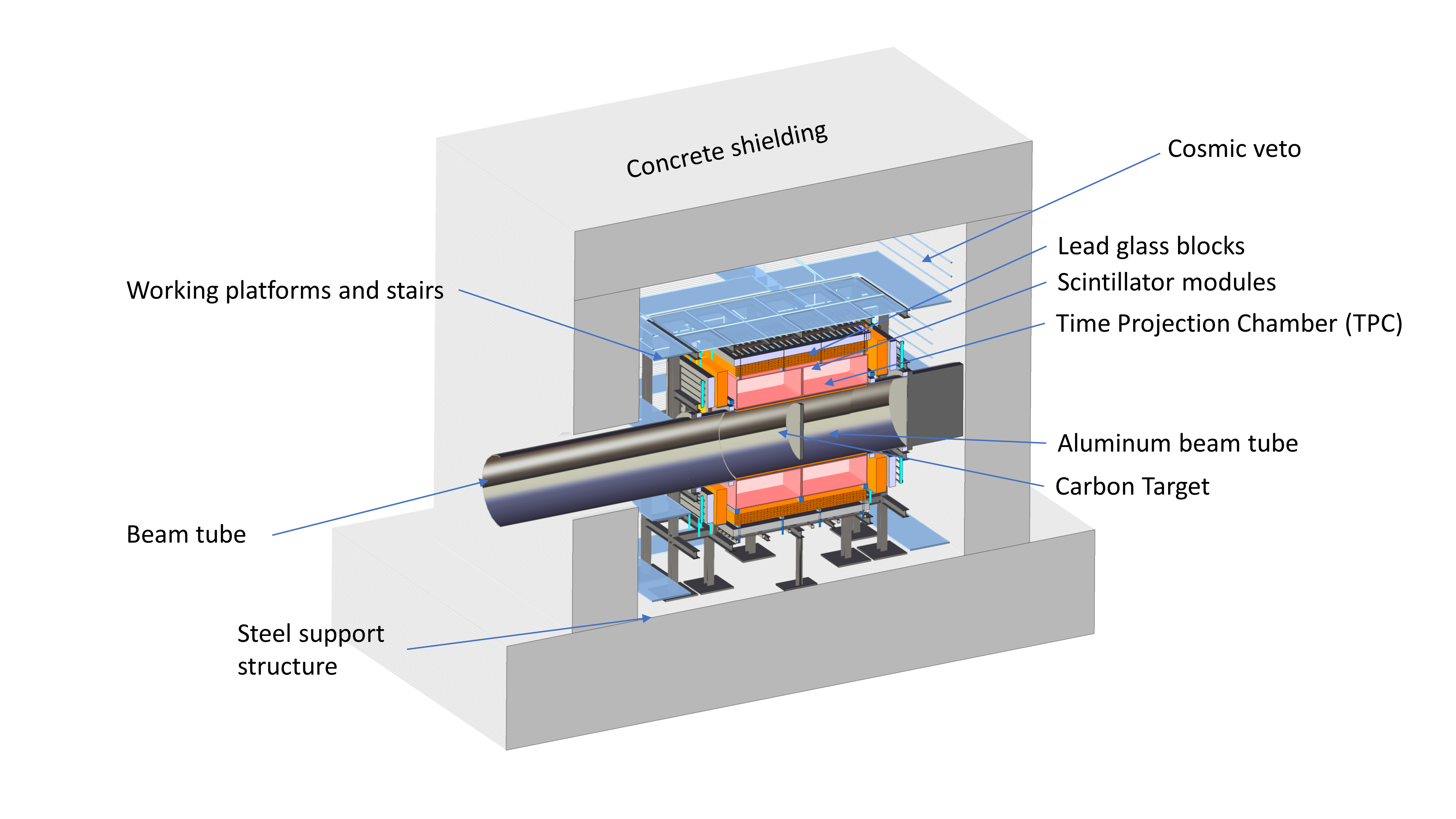}
    \caption{Cut-away of the NNBAR annihilation detector from Ref.~\cite{Santoro:2024lvc}.}
    \label{fig:detp}
  \end{subfigure}
 \caption{A schematic visualization of the exotic neutron decay experiment setup (a) along with a cut-away CAD drawing of the NNBAR detector (b).}
  \label{fig:combined_figures}
\end{figure}

\textcolor{black}{The concept is illustrated in Figure~\ref{fig:combined_figures}, which shows (a) a schematic visualization of an exotic neutron decay experiment, where a flux of cold neutrons passes through a beam pipe enclosed within a detector volume, and (b) a cut-away CAD drawing of the proposed NNBAR annihilation detector. In this setup, any rare neutron decay occurring within the central beam pipe---which is approximately 6.5\,m in length---would be detected by a system composed of hadronic and electromagnetic calorimeters, as well as a particle tracker. The full detector volume spans approximately $5 \times 5\,\mathrm{m}^2$ in transverse dimensions. The primary background for this configuration originates from cosmic-ray particles.}

Since the final-state particles would typically possess kinetic energies exceeding 300~MeV, accurately measuring this energy for charged pions or muons is challenging due to calorimeter punch-through. However, it would still be possible to determine if the deposited energy is significant. 
Additionally, tracking would allow the inference that the final-state particles are back-to-back, and time-of-flight measurements would further aid in suppressing cosmic-ray backgrounds.

Final states involving muons are particularly susceptible to cosmic-ray backgrounds and are therefore less favorable from an experimental standpoint. Channels involving electrons and gammas, on the other hand, are excellent candidates since their energy can be precisely measured with the NNBAR and HIBEAM annihilation detectors. Specifically, channels with high-energy photons in the final state, including those from neutral pion decays, are promising since the electromagnetic calorimeter can measure their energy with high accuracy. The channel \( X \rightarrow \nu + \gamma \) would appear as an imbalanced high-energy gamma in the detector, making it a well-measurable channel.

Since searches for exotic neutron decays will be conducted parasitically alongside neutron-antineutron searches, it will be crucial to exclude any events originating from the thin carbon target that is part of the annihilation detector.

Detector efficiencies are expected to be at least 50\%, which—based on prior experience designing the NNBAR detector for neutron-to-antineutron searches~\cite{Yiu:2022faw}—is considered a conservative estimate. Previous analyses prioritized maximizing signal efficiency while ensuring that no background events survive selection, a strategy enabled by a deep understanding of detector response and background rejection techniques. While updated Geant4 simulations~\cite{GEANT4:2002zbu} will be necessary to quantitatively confirm background-free selection for new topologies, particle identification and core rejection strategies are expected to remain mostly unchanged.

\textcolor{black}{This zero-background assumption extends naturally to exotic neutron decay channels. Although the original cuts \cite{Santoro:2024lvc} were optimized for multi-pion final states, the key tools—timing-based calorimeter hit selection, filtered energy loss in lead-glass modules, object multiplicity, and topological observables—are largely topology-independent and effective against cosmic-ray backgrounds. For channels proposed for the HIBEAM detector, such as \( X \to \nu + \gamma \) and \( X \to \pi^0 + \gamma \), these techniques remain applicable. In \( X \to \nu + \gamma \), the signal consists of a single isolated photon without associated charged tracks, allowing for efficient suppression via timing cuts, Time Projection Chamber (TPC) track vetoing, and an active cosmic veto. In \( X \to \pi^0 + \gamma \), photon pair reconstruction enables \(\pi^0\) identification, adding discriminating power on top of timing and energy criteria. While HIBEAM differs from NNBAR in having a smaller transverse size and lacking internal silicon tracking, its calorimetric and TPC capabilities preserve the essential handles for background suppression.} 

\section{Conclusion}
\label{conclusion}
In this work, we have explored the possibility of searching for visible high-energy products resulting from exotic neutron decays. We have demonstrated that this idea is not ruled out by large-volume detectors. Furthermore, no previous free-neutron experiment has been capable of investigating this possibility, making the HIBEAM-NNBAR experimental program at the European Spallation Source the best opportunity for such a test.

Even at its highest flux stage, NNBAR, the reach of the experiment will be constrained to scenarios with a high degree of degeneracy with the neutron, due to the requirement of a long detector for measuring the decays. However, this remains the most theoretically motivated scenario, as such fine-tuning would suggest the presence of a new symmetry, which naturally leads to nearly degenerate masses.

While this work is speculative in nature, it is nonetheless worthwhile to verify. No additional experimental setups are required, as the planned HIBEAM-NNBAR infrastructure is sufficient to carry out the test in a fully parasitic manner, requiring only offline analysis. Further improvements in the precision of neutron branching ratio measurements would impact our conclusions only by reducing the available parameter space for such exotic decays in free-neutron experiments, but they would not rule them out entirely. Given that particle degeneracy is a recurring theme in nature, we believe that an effort to perform the necessary analysis is well justified.

\ack
We thank Anders Oskarsson for valuable discussions and insights regarding the experimental setup of this work. We thank Richard Wagner for his assistance in estimating the NNBAR neutron flux. 
We are especially grateful to Antonio Rodríguez-Sánchez for his valuable suggestions, corrections, critiques, and for reviewing some of our calculations.
We are truly thankful for valuable comments by José Zurita on the manuscript.

\section*{References}
\bibliographystyle{unsrt}
\bibliography{references}

\end{document}